# Designing Actively Secure, Highly Available Industrial Automation Applications

Awais Tanveer, Roopak Sinha and Stephen G. MacDonell
*IT & Software Engineering, Auckland University of Technology, Auckland, New Zealand*
Email: awais.tanveer@aut.ac.nz

Paulo Leitao
*Research Centre in Digitalisation and Intelligent Robotics (CeDRI), Instituto Politecnico de Braganca, Campus de Santa Apolonia, Braganca, Portugal*
Email: pleitao@ipb.pt

Valeriy Vyatkin
*Aalto University, Helsinki, Finland Luleå University of Technology, Sweden*
Email: vyatkin@ieee.org

**Abstract**

*Programmable Logic Controllers (PLCs) execute critical control software that drives Industrial Automation and Control Systems (IACS). PLCs can become easy targets for cyber- adversaries as they are resource-constrained and are usually built using legacy, less-capable security measures. Security attacks can significantly affect system availability, which is an essential requirement for IACS. We propose a method to make PLC applications more security-aware. Based on the well-known IEC 61499 function blocks standard for developing IACS software, our method allows designers to annotate critical parts of an application during design time. On deployment, these parts of the application are automatically secured using appropriate security mechanisms to detect and prevent attacks. We present a summary of availability attacks on distributed IACS applications that can be mitigated by our proposed method. Security mechanisms are achieved using IEC 61499 Service-Interface Function Blocks (SIFBs) embedding Intrusion Detection and Prevention System (IDPS), added to the application at compile time. This method is more amenable to providing active security protection from attacks on previously unknown (zero-day) vulnerabilities. We test our solution on an IEC 61499 application executing on Wago PFC200 PLCs. Experiments show that we can successfully log and prevent attacks at the application level as well as help the application to gracefully degrade into safe mode, subsequently improving availability.*

## 1. INTRODUCTION

The fourth industrial revolution commonly knowns as Industry 4.0, critically relies on networked automation systems with decentralized decision making and connectivity beyond the confines of the factory. Such an exposed posture attracts cyber adversaries that may result in devastating effects if critical industrial processes get disrupted from their normal behavior. Industrial Automation and Control Systems (IACS) enable the automation and control of physical processes right from the factory floor to higher level data analytics at the enter- prise level. Communication between SCADA components and the outside world forms a heterogeneous network comprising of different technologies and protocols. The heterogeneity adds to the complexity of overall system design that may limit protection against adversaries. Consequently, establishing robust security solutions becomes a paramount requirement for SCADA systems.

PLCs are critical components of IACS. They control physical processes and provide floor data to the upper layers for processing. PLCs read data from multiple sensors, processes the data through installed software applications, and then either send messages to other PLCs or drive actuators. A secure IACS needs robust PLC-level protection from attacks on peripherals and networks. In the past, PLC-level security attacks, such as Stuxnet [1], have posed significant threats to critical infrastructure. An analysis of PLC protocols such as UMAS, S7Comm, and Optocomm-Forth on Schneider PLCs revealed several vulnerabilities like user program comprise and alteration, configuration compromise, authentication/access control violation, etc. [2]. In another instance, researchers successfully carried out the replay, man-in-the-middle and stealth command modification attack on PLC devices [3]. Successful Denial of Service (DoS) attacks were carried out on real PLC devices rendering them unresponsive [4].

IACS are meant to run for long durations. It is desirable that the software controlling physical processes in an IACS remains available with minimum downtime, which makes system *availability* a principal goal. Availability attacks, such as DoS-type attacks, on even a single PLC can sabotage the entire IACS [4]. In fact, for IACS, availability is a more critical security concern than confidentiality and integrity [5]. PLCs are vulnerable to external availability attacks as well as internal attacks [4], [6]. The resource-constrained nature of PLCs makes it harder to provide absolute protection against sophisticated external and internal attacks. Sometimes, an Intrusion Detection and Prevention System (IDPS) may be used to detect and prevent attacks against PLCs [7]. An IDPS continuously monitors traffic over a network and/or the internal file system or memory footprint of its host. If anomalous or unexpected patterns are detected, it may take preventative actions to secure the PLC. IDPS usually acts as a last line of defense for PLCs in a network. However, even the most sophisticated IDPS may also fail to detect an attack, especially *zero-day* vulnerabilities that are attacked for the first time [4].

In this paper, we investigate the case for providing security protection at the application level of a PLC, such that attacks that circumvent an IDPS can be intercepted by the software application running on the PLC. Application-level security needs to be light-weight as PLCs are highly resource con- strained. However, this additional layer of security can provide active security protection against zero-day attacks, helping to improve overall system availability. We first study the most commonly-encountered availability attacks on PLCs. For the mitigation, we use the IEC 61499 Service Interface Function Blocks (SIFB) that implements IDPS-like functionality within applications. IEC 61499 provides a highly modular application structure where large applications can be constructed through the reuse of much smaller and pre-verified function blocks from a library. We implement and evaluate a proof-of-concept distributed IEC 61499 function blocks application running over Wago PFC200 PLCs. The security functions are provided through SIFB executing Snort, a well-known IDPS [8]. We emulate availability related attacks on the proof-of-concept application through multiple tools. Subsequently, performance analysis of Snort in terms of packet dropped over a time interval, running in IEC 61499 environment, is provided. The results show that the Snort drops more packets with the increasing intensity of attacks losing critical data for analysis purposes. At a certain intensity, the PLC device also breaks down, resulting in total denial of service.

The primary contributions of this paper are:

1. A summary of commonly-encountered availability at- tacks on PLC applications in IEC 61499 environment, presented in Section II.
2. The design of an IDPS-based SIFB for preventing avail- ability attacks, described in Section III.
3. A sample implementation of a secure, distributed IEC61499 application and its evaluation, presented in Section IV.

## 2. AVAILABILITY ATTACKS ON IEC 61499 APPLICATIONS

This section presents the scenarios for IEC 61499 dis- tributed applications in which availability attacks might disrupt the operations of the system. We list the most common and easily replicable availability attack vectors, mapped to IEC 61499 IACS applications. A hypothesis is formed for each attack scenario that is tested in the subsequent section.

To map availability attacks on a real example, consider a scenario where two cylinders controlled by two cooperating PLCs as seen in Fig. 1. In Fig. 1 (a), both cylinders are initially at retracted state while `cylinder 2` is waiting for the box to arrive at its plate. As soon as the box arrives at `cylinder 2`, the controlling PLC detects it through a sensor and commands the cylinder to start extending. When the box reaches at the top, as seen in Fig. 1 (b), the PLC controlling `cylinder 1` signals it to extended until it pushes the box off the ledge of `cylinder 2`. At this point, both cylinders are at the fully extended position. To get back to the original position, the PLC controlling `cylinder 1` issues a signal to the PLC controlling `cylinder 2` to retract as seen in Fig. 1 (c).

Fig. 2 shows the IEC 61499 implementation of case study described in Fig. 1. The PLC that controls cylinder 1 hosts `ThrustCtl` and IX Function Block (FB)s. Similarly `LiftCtl` and QX FBs reside on the other PLC that actuates cylinder 2. The `ThrustCtl` FB extends the cylinder on receiving the signal from PLC through IX FB. When fully extended, the `ThrustCtl` FB sends the `SharedVariable` to `LiftCtl` on the other PLC while retracting the `cylinder 1`. On receiving the `SharedVariable` data input, `cylinder 2` starts to retract as well.

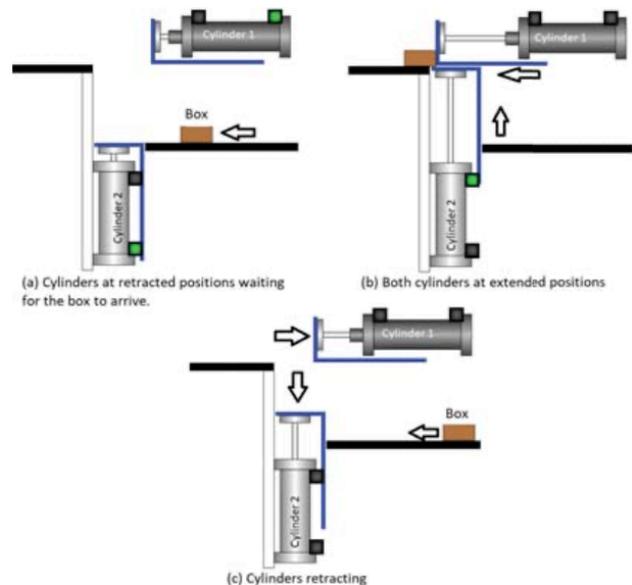

Fig. 1. Illustration of scenario in which the proposed SIFB has been implemented.

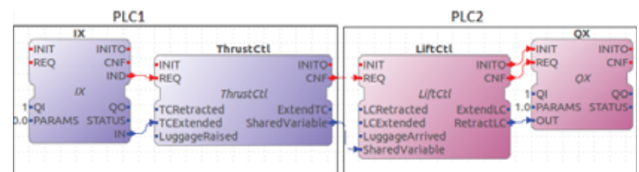

Fig. 2. IEC 61499 implementation of case study in fig. 1

### A. Attack with malicious or malformed data
The application in Fig. 2 (distributed amongst PLC1 and PLC2) transfers the `SharedVariable` using publisher/subscriber model using UDP multicast network.

We assume that there is no confidentiality or integrity support for standard or propriety communication protocol used between the publisher and subscriber. The situation is true for legacy PLCs operating in small to medium-sized industrial setups [9]. Multicast and unicast modes of communication are defined in the HOLOBLOC profile for IEC 61499 standard. In this mode, any host can become a part of a multicast group and send messages to the participants of the group. We construct the following hypothesis for this scenario:

*Hypothesis 1 (H1):* An adversary can send malicious data to the subscriber or server block by masquerading itself as publisher/client Communication Service Interface Block (CSIFB), causing it to misbehave and subsequent disruption of the intended service.

*Assumption(s):*
1) There is no mechanism in place for confidentiality or integrity checks.
2) An adversary can send data to the participants in the multicast group.

The attack model is valid for unicast mode as well. IEC 61499 client/server CSIFBs use the TCP mode of communication. Although spoofing a packet is harder in TCP than UDP, an attacker may apply a replay technique to spoof a TCP packet to hijack the session. In this case, `LiftCtl` FB will act upon the instructions sent in the packet by the attacker instead of the legitimate client.

The `ThrustCtl` FB in PLC1 is sending periodic data from a sensor to the instance in PLC2 that is controlling an actuator based on the received information. The data format can be as simple as presenting integers, in this case, a Boolean variable i.e., `SharedVariable`. An adversary may be able to intrude in the system and may realize the importance of periodic messaging by observing the behavior of the system. It can take advantage of non-existent security controls and may send malicious data to `LiftCtl` in PLC2 using packet crafting techniques by masquerading as PLC1. The subscriber or server block will receive out of sync messages to act on actuators that will cause the system to misbehave or may even cause permanent damage to the machinery. Fig. 3 (a) illustrates the discussed scenario where an attacker sends a fake packet to `LiftCtl` FB containing `SharedVariable`, forcing uncoordinated actuation from both PLCs.

**B. Application level flood attack**

Another scenario is the possibility of DoS as well as Distributed DoS (DDoS) attack where multiple attackers can choke the system to stop or delay the response of service. Fig. 3 (b) depicts a situation where it is possible to suffocate `LiftCtl` and `ThrustCtl` FBs by sending the flood of malicious traffic from multiple attackers. DDoS attacks are difficult to manage if there are no appropriate security controls implemented in the boundary network. In this case, the communication between publisher and subscriber FBs will be rendered non-functional by the adversaries participating in the DDoS attack as the legitimate traffic from the publisher will not able to reach the subscriber.

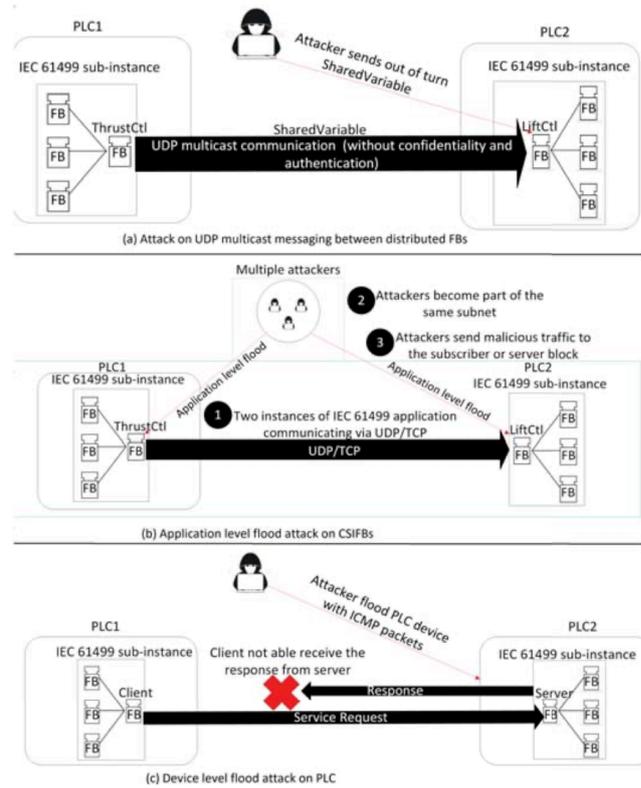

Fig. 3. A subset of attack scenarios of IEC 61499 distributed applications

*Hypothesis 2 (H2):* One or multiple adversaries can become a part of the multicast group and flood the publisher/subscriber interface to make it unavailable or slow to respond to legitimate traffic.

*Assumption(s):*
1) An adversary can send data to the participants in the multicast group or TCP port.
2) An adversary has been able to penetrate through the network firewall.

Again, in the case of an application using client/server CSIFBs, an attacker can flood the port bound by the server CSIFB with invalid requests. TCP attacks such as SYN flood attack is particularly famous for causing a denial of service for TCP sockets. In this attack, repeated SYN packets are sent to the victim at a high rate. The victim will send SYN-ACK packets back to the attacker and wait for the ACK that never arrives, thus leaving the TCP session in a half-open state. If a large number of half-open connections reach a tipping point, the victim will start denying legitimate connection requests. Thus, a server CSIFB with a TCP socket is also vulnerable to such an attack.

The repercussions can be noticed if an attacker manages to flood the TCP/IP ports bound by `LiftCtl` and `ThrustCtl`. The application-level flood attack may overwhelm the FB interfaces with legitimate traffic. Such a scenario may incur no or out-of-sync communication between PLC1 and PLC2 that can halt the system or even inflict irreparable damage to the equipment. Such attacks on PLCs are not unheard of, at least in smart grids where [10] has shown a model of power systems with DoS attacks. It shows that communication channels leading to the load frequency controller of a power system can be jammed, resulting in the loss of telemetered measurements.

**C. Device level flood attack**

A PLC may become vulnerable to flood attacks at device level such as ICMP or ping flooding. Such a type of attack inundates the device communicating through TCP/IP stack with brute force i.e., sending a large number of ICMP packets at a rapid rate by a single or multiple attackers. Inadvertently, the device starts consuming its resources to process these packets, consequently denying the resources to legitimate traffic.

Although, the problem can be considered in isolation to IEC 61499 applications as it affects the whole system. However, denial of services provided by an instance of distributed IEC 61499 application will eventually affect the other instances running on other PLCs. The distributed nature of IEC 61499 applications means that the instances will be partially or entirely dependent on each other. Unavailability of the services an FB under attack may cause the dependent FB unable to perform a critical system functionality unless mechanisms are put in place in the design of a distributed application by foreseeing DoS attacks. Fig. 3(c) represents a scenario where a PLC device acting as a server, may be swamped by ICMP flood attack so that it stops responding to legitimate requests by a client device. Therefore, the security of the IEC 61499 distributed environment against device-level DoS attacks may not be considered as a stand-alone problem. Corresponding to the case study in Fig. 1, a device level flood attack on either PLC1 or PLC2 may overwhelm the PLCs slowing down the response time or may lead to an unresponsive state.

*Hypothesis 3 (H3):* One or multiple adversaries can flood the PLC running an instance of IEC 61499 distributed application to make it unavailable for other dependent instances.

*Assumption(s):*
1) An adversary knows the IP address of the PLC device.
2) An adversary has been able to penetrate through the network firewall.

## 3. AN SIFB BASED IDPS

Application and device level attacks can be prevented by a firewall. A firewall's main goal is to block illegitimate traffic based on a packet's header information. On the other hand, an IDPS is also able to analyze the packet's data to raise an alert. IDPS has a distinct advantage over a firewall because it can detect the abnormal behavior of an attacker once it gets in. Thus, an IDPS can essentially be considered as the last line of defense to secure a communication system. Fig. 4 shows a generic scenario of a network configured to have an IDPS.

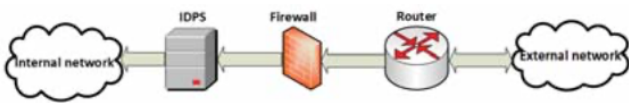

Fig. 4. A simple network configuration containing IDPS

In small to medium-sized industrial settings, a dedicated IDPS hardware solution is not always cost-effective. The absence of intrusion detection controls puts such environments at high risk. However, IDPS can be deployed in a PLC device, as demonstrated in [11]. The modular structure of IEC 61499 FBs allows the provision of deploying an IDPS module as an FB. In this section, we propose an IEC 61499 distributed application model incorporating intrusion detection and prevention capabilities. Moreover, we also discuss the interface design and utilization of SIFB embedding IDPS services. IEC 61499 standard allows distributed instances of an application to communicate through publisher/subscriber or client/server CSIFBs that put or acquire the data to or from the network. Fig. 5 illustrates our proposed model of an IEC 61499 application comprising of an IDPS. A CSIFB provides an interface to the underlying network. It then transfers the data to or from the FBs providing the application's business logic. In the case of Fig. 5, the CSIFB is configured to receive an input from the network and pass it to the back-end logic component. Therefore, the back-end component is dependent on the CSIFB. The proposed model places an IDPS in between network interface and logic components so that it acts as a Boolean switch. That is, if the IDPS component detects an attack, it shuts off the connection between CISFB and logic components until the attack is on. The designer may decide to suspend the application instance or completely shut down the process depending on the requirements.

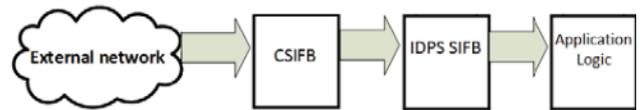

Fig. 5. Proposed configuration of IDPS SIFB in IEC 61499 distributed applications

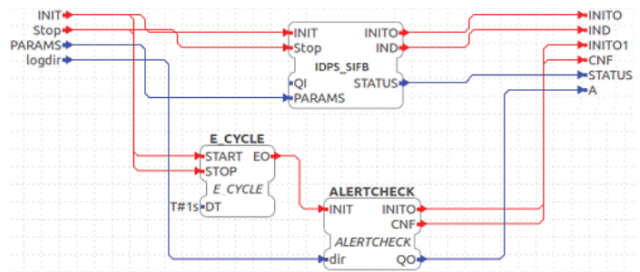

Fig. 6. A CFB containing `IDPS_SIFB` and associated `ALERTCHECK` FB.

The model can essentially be realized by creating a SIFB that starts the IDPS process in the background upon receiving the `INIT` event. The `stop` event is used to stop the IDPS process. The `PARAMS` data input is used to provide the necessary configuration parameters to the process. The `STATUS` data output can be used to check the state of the process. The `IDPS_SIFB` is essentially an integral part of a composite FB (CFB) called `IDPS_CFB`. Fig. 6 shows the internal FB network of `IDPS_CFB`. It includes `IDPS_SIFB` and an `ALERTCHECK` FB which continuously checks the output of `IDPS_SIFB` to determine the occurrence of an attack by looking for alerts generated by an IDPS. If the attack is on, `ALERTCHECK` FB will raise a Boolean flag through `QO` variable that is connected to A data output of the CFB. The value can subsequently be consumed by the encapsulating application to enable the appropriate action if a PLC device or an application is under attack.

The most important point of reference in the system is the extended position of both the cylinders. At this stage, the `ThrustCtl` FB causes the `cylinder 1` to retract, also switching on the `SharedVariable` Boolean variable and sends it to `LiftCtl` controlling cylinder 2 on the other PLC. On the receiving `SharedVariable`, the `LiftCtl` FB signals the PLC to retract cylinder 2. Therefore, the only point of interaction between two PLCs is the exchange of `SharedVariable`. In the case when Ethernet connects the PLCs, the data travels in TCP or UDP packets depending on the mode selected i.e., client/server or publisher/subscriber model. In the case study, the communication between two PLCs controlling the cylinders is vulnerable to the availability attacks presented in section II. For example, an attacker can interrupt the transmission of `SharedVariable` by rendering one of the PLC unresponsive through a DoS or DDoS attack. In that case, the congruity of the system shall be affected that may lead to irrecoverable physical damage to the equipment in focus.

However, the availability of a mechanism that can detect such attacks may enable designers and developers of an IEC 61499 to pre-empt against such attacks. In the case of IDPS, the attack can be prevented and blocked so that the system may continue to perform its routine task. On the other hand, an Intrusion Detection System (IDS) can report the occurrence of an attack, prompting the system to go into a defensive posture e.g., halting the system for a certain amount of time or any discretionary actions. A possible solution is provided in Fig. 7 to establish the application of the aforementioned `IDPS_CFB`. In an event of availability attack on the PLC controlling `cylinder 2`, the `IDPS_CFB` is able to report on the attack. If the Boolean variable A is true, the `LiftCtl` FB is not able to receive the shared variable resulting in halting the process. Such a mechanism is useful if an out of sync `SharedVariable` input arrives due to an ongoing attack.

Moreover, the optional `E_SWITCH` FB in Fig. 7 connects the `IDPS_CFB` to the application logic. In intrusion prevention mode, it performs an important job to switch off the application logic during the occurrence of an attack. However, in the intrusion prevention mode, when `IDPS_CFB` is actively blocking the malicious traffic, the `E_SWITCH` FB can be removed to create a direct connection between IDPS and application logic. It will enable the designer to log the defined attacks while the application is performing its regular tasks. It will positively affect the availability in a way that the application will have minimal downtime and also allows mitigation measures simultaneously.

**Active security protection:** An important factor in protecting against availability related attacks is to put the system into a security posture that is active/proactive rather than reactive. Although organizations are investing more in security infrastructure as IACS are becoming more social, attackers are also evolving armed with sophisticated attack techniques. The signature or rule-based IDPSs, although convenient, can- not detect an innovative attack that does not make part of their ruleset or signatures. A possible proactive approach is to use anomaly-based IDPS using Machine Learning (ML) techniques to extract useful and harmful patterns. Thus a model can be obtained for correct system behavior. Alarms are raised when such an IDPS detects anything that defies the model. However, ML techniques usually require more processing power that is a limitation for PLC devices in IACS. Therefore, lightweight ML algorithms may be chosen to get the best of both worlds in such devices. An advantage of using the ML-based approach, striving to provide active security protection is its suitability for detecting zero-day attacks to some extent. The method to use rule-based Snort IDPS with ML algorithms has been discussed in [12]. It uses ML-based intelligent plugin along with Snort to classify good and bad traffic. The ML plugin runs parallel to Snort's detection engine with the ability to set alarms for newly identified attacks. However, a possible extension to this approach will be to feed newly detected attacks back to Snort in the form of new rule sets. A proactive security posture can be achieved for IEC 61499 environment by using the proposed IDPS_SIFB using lightweight ML algorithms.

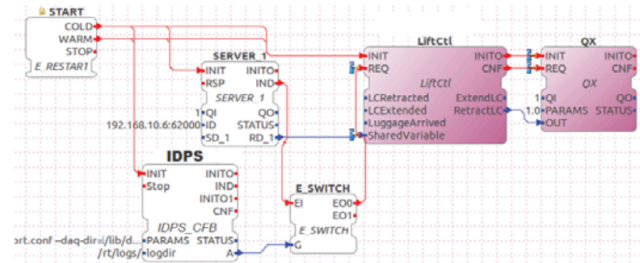

Fig. 7. Application of `IDPS_CFB` with `LiftCtl` FB.

## 4. IMPLEMENTATION AND RESULTS

We have partially implemented the case study described in section II. Two separate Wago 750-8206 PFC200 PLCs control both cylinders. Each PLC has an ARM one core Cortex A8 600 MHz microprocessor with 256 MB main memory running real-time Linux. The Wago PLC supports IEC 61499, which was used to develop the application. The PLCs communicate over Ethernet connected via a switch. We used the 4DIAC IDE and FORTE [13] to develop the application logic, and then cross-compiled it to run on realtime Linux in the Wago PLCs. Sensor input to PLC1 indicates that `cylinder 1` is fully extended. In response, it sends the `SharedVariable` signal to PLC2. Consequently, PLC2 commands `cylinder 2` to retract in the form of an actuator signal. In our set up, we used a metallic proximity sensor as the sensor input to PLC1. Fig. 8 shows the experimental set up used to imitate the partial system behavior.

We have used the Snort IDPS [8] to implement IDPS_CFB, the block that implements the IDPS at the application level. Snort can run in IDS as well as IDPS modes, but we have so far only implemented the IDS mode in IDPS_CFB. We emulate DoS attacks using the hping3 tool [14] which sends a large number of ICMP and TCP packets to PLC1. The INIT event of IDPS_CFB starts a new Snort process externally to the FORTE runtime. Snort rules are configured in a configuration file to detect availability attacks. Such rules enable Snort to detect any unwanted traffic and raise alerts or block traffic in IDPS mode. As Fig. 9 shows, `SnortCFB` sets variable A when it detects traffic violating the rules.

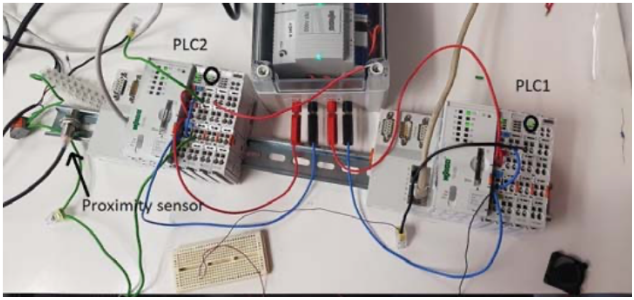

Fig. 8. Experimental set up with Wago 750-8206 PFC200.

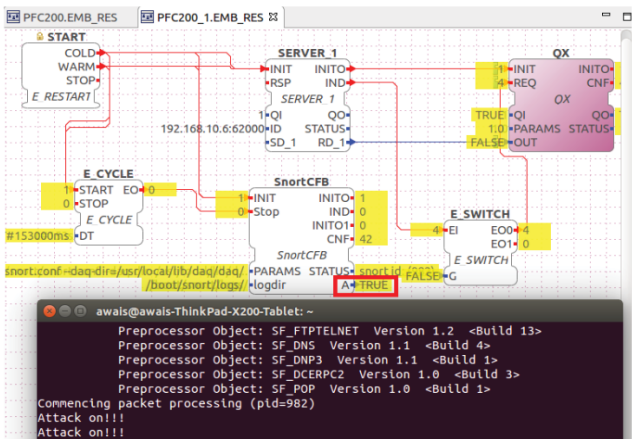

Fig. 9. Snort based `IDPS_CFB` detecting attacks.

To prove hypothesis *H1* in section II, the PackETH [15] tool is used to send a spoofed UDP packet to cylinder 2 containing the value of `SharedVariable`. It causes out-of- sync actuation of `cylinder 2` at PLC2. For hypotheses *H2* and *H3*, hping3 is used to send a large number of packets to the PLC host and its ports used by the FORTE runtime.

We observed the number of packets dropped by Snort as the packet frequency increased. In high-intensity attacks, Snort fails to log the attack correctly. When *hping3* was configured with the *–faster* option to send packets each microsecond, the PLC itself becomes completely unresponsive which halts the operation of the system. It shows that a sufficiently powerful attacker can succeed even in the presence of an IDPS system embedded in the IEC 61499 application. However, the application-level IDPS can be very useful in logging and/or filtering out illegitimate traffic that escapes other mitigation strategies during low to medium intensity attacks.

The demonstrated behavior of IDPS shows the viability and applicability of PLC-based solutions in industrial settings that may improve the overall availability of the PLC and thus the system under consideration. The IDPS_CFB is essentially the last resistance in the path of an attacker if it can breach upper layers of an IACS. The ideal choice in such a setup is an underlying lightweight IDPS to serve the balance between security and other properties of the system that are affected by the PLC such as safety and efficiency. However, identifying the appropriate IDPS suitable to PLCs deployed in different environments is a constraint on the solution provided in the research.

## 5. CONCLUSIONS AND FUTURE WORK

In this paper, we have enumerated the most common attack vectors that may affect the availability of distributed industrial applications. We propose an IEC 61499 SIFB based IDPS solution to protect PLCs against such availability attacks. An implementation A prototype application was built to implement this solution using Snort IDPS. Results show that under low to medium attacks, the IDPS can successfully log illegitimate traffic, providing the last line of defense against attackers. This work is a first step towards the use of more sophisticated detection and prevention algorithms to provide active security protection at the application level. Future research directions include formalizing the solution as a replicable design pattern in IEC 61499, creation of a SIFB library for different detection and prevention algorithms, and the design of high availability IACS through system-, PLC-, and application-level mitigation. Moreover, we also plan to carry out the detailed comparison, evaluation, and benchmarking of `IDPS_CFB` using multiple off-the-shelve IDPSs in the future.